\documentclass[aps,physrev,reprint,longbibliography,groupedaddress,floatfix]{revtex4-2}

\usepackage{graphicx}
\usepackage{amsmath}
\usepackage{amssymb}
\usepackage{bm}
\usepackage[hidelinks]{hyperref}

\begin{document}

\title{An Accelerator-Based Vacuum-Ultraviolet Frequency Comb for Thorium-229 Nuclear Spectroscopy}

\author{Junhao Liu}
\affiliation{Shanghai Advanced Research Institute, Chinese Academy of Sciences, Shanghai 201210, China}
\affiliation{University of Chinese Academy of Sciences, Beijing 100049, China}

\author{Zhen Wang}
\affiliation{Shanghai Advanced Research Institute, Chinese Academy of Sciences, Shanghai 201210, China}
\affiliation{University of Chinese Academy of Sciences, Beijing 100049, China}

\author{Tao Liu}
\affiliation{Shanghai Advanced Research Institute, Chinese Academy of Sciences, Shanghai 201210, China}
\affiliation{University of Chinese Academy of Sciences, Beijing 100049, China}

\author{Chao Feng}
\email{fengc@sari.ac.cn}
\affiliation{Shanghai Advanced Research Institute, Chinese Academy of Sciences, Shanghai 201210, China}
\affiliation{University of Chinese Academy of Sciences, Beijing 100049, China}

\author{Zhentang Zhao}
\email{zhaozt@sari.ac.cn}
\affiliation{Shanghai Advanced Research Institute, Chinese Academy of Sciences, Shanghai 201210, China}
\affiliation{University of Chinese Academy of Sciences, Beijing 100049, China}

\begin{abstract}
Vacuum-ultraviolet frequency combs with high power per line could combine
high-resolution spectroscopy with coherent control of quantum transitions.
Here we propose using a linear accelerator to generate such a comb for
$^{229}$Th nuclear spectroscopy.
Radiation simulations show that precompensating the electron
microbunching phase narrows the single-pulse spectrum and reduces
shot-to-shot phase fluctuations at a nuclear resonance. Nuclear models
show how the high power in individual comb teeth can support solid-state
nuclear-clock spectroscopy and drive nuclear Rabi oscillations in trapped
ions. In the ion model, two adjacent teeth of the same comb prepare a
nuclear superposition whose relative phase can be read out through
Ramsey interference.

\end{abstract}

\maketitle

\section{Introduction}

Optical atomic clocks use electronic transitions as frequency references.
Nuclear transitions can offer reduced sensitivity to external fields, making
them attractive for precision metrology~\cite{Campbell2012}. The unusually
low-lying isomer in $^{229}$Th places a nuclear transition within reach of
vacuum-ultraviolet (VUV) lasers near $148.4\,\mathrm{nm}$. Direct laser
excitation was demonstrated in thorium-doped
crystals~\cite{Tiedau2024,Elwell2024}. Subsequent VUV frequency-comb
spectroscopy resolved the nuclear transition lines split by the crystal
electric field gradient and measured the transition frequency relative
to an $^{87}$Sr optical clock~\cite{Zhang2024}.

Narrow-linewidth continuous-wave (CW) VUV sources are now
available~\cite{XiaoCW2026}. Experiments have demonstrated direct nuclear
absorption spectroscopy~\cite{Morawetz2026} and closed-loop operation of
solid-state $^{229}$Th clocks~\cite{Huang2026,ToscaniDeCol2026}. VUV
frequency combs connect nuclear structure measurements to optical
frequency metrology by resolving individual transitions across a broad
spectrum on a common scale traceable to an optical clock~\cite{Zhang2024}. Increasing the power in a resonant tooth can
shorten delayed-fluorescence acquisition at a given signal-to-noise ratio
(SNR) [Sec.~\ref{sec:nuclear_spectroscopy}]. When both are on resonance, two
mutually coherent teeth can drive a pair of transitions with a defined
relative phase. Two-field quantum interference has been modeled in
$^{229}$Th-doped crystals~\cite{Das2013}.

In trapped $^{229}$Th$^{3+}$ ions, electronic hyperfine spectroscopy has
distinguished the nuclear ground and isomeric states~\cite{Yamaguchi2024}.
Observing nuclear Rabi oscillations driven directly by VUV light in a single
trapped ion remains an experimental goal~\cite{Moritz2025}. Resolving these
oscillations calls for sufficient resonant VUV power at the ion to drive
substantial population transfer within the coherence
time~\cite{wense2020,Xiao2026}.

Superconducting radio-frequency linear accelerators (linacs) can supply
high-repetition-rate electron bunches for high-average-power undulator
radiation~\cite{Decking2020}. In an
externally seeded undulator source, the seed sets the phase reference for
the harmonic microbunching, and the electron beam supplies the radiated
energy~\cite{Allaria2012}. If
successive pulses remain mutually phase coherent, the bunch repetition rate
sets the comb-tooth spacing. For the same normalized single-pulse spectrum
spanning many teeth, wider spacing places a larger fraction of the average
power in the tooth aligned with the nuclear transition.

Modulating every electron bunch at gigahertz repetition rates places a
demanding average-power requirement on the seed laser.
Angular-dispersion-induced microbunching (ADM) uses multidimensional beam
manipulation to reduce the suppression of bunching by uncorrelated
energy spread, allowing strong harmonic density modulation with weak energy
modulation~\cite{FengZhao2017,Liu2026}. One design study estimated
that ADM requires about three orders of magnitude less seed peak power
than conventional harmonic generation~\cite{Zhao2021}.

Here we propose a $1.3\,\mathrm{GHz}$ accelerator-based ADM VUV frequency comb for
$^{229}$Th nuclear spectroscopy. We introduce microbunching-phase
precompensation (MPP), using dogleg dispersion to offset the phase
curvature acquired in the radiator. Three-dimensional simulations show
that MPP raises the peak spectral energy density and, under combined
beam and seed-power jitter, improves shot-to-shot coherence in the
selected transverse mode at the target transition. At the radiator
exit, the simulated field gives a sub-meV single-pulse spectral full
width at half maximum and, for an ideal coherent pulse train,
$21.8\,\mathrm{mW}$ in a tooth aligned with a crystal nuclear transition.
Nuclear models predict faster fluorescence readout of a solid-state
clock transition and, with milliwatt-level power delivered to a trapped
ion in one resonant tooth, submillisecond $\pi$ pulses for nuclear Rabi
driving. We also propose preparing a nuclear superposition with two
adjacent teeth of the same comb and reading out its relative phase from
the modeled Ramsey fringe in the ground-state return probability. Such
control and readout could support quantum memories based on the
long-lived $^{229}$Th isomer.

\begin{figure*}[t]
\centering
\includegraphics[width=\textwidth]{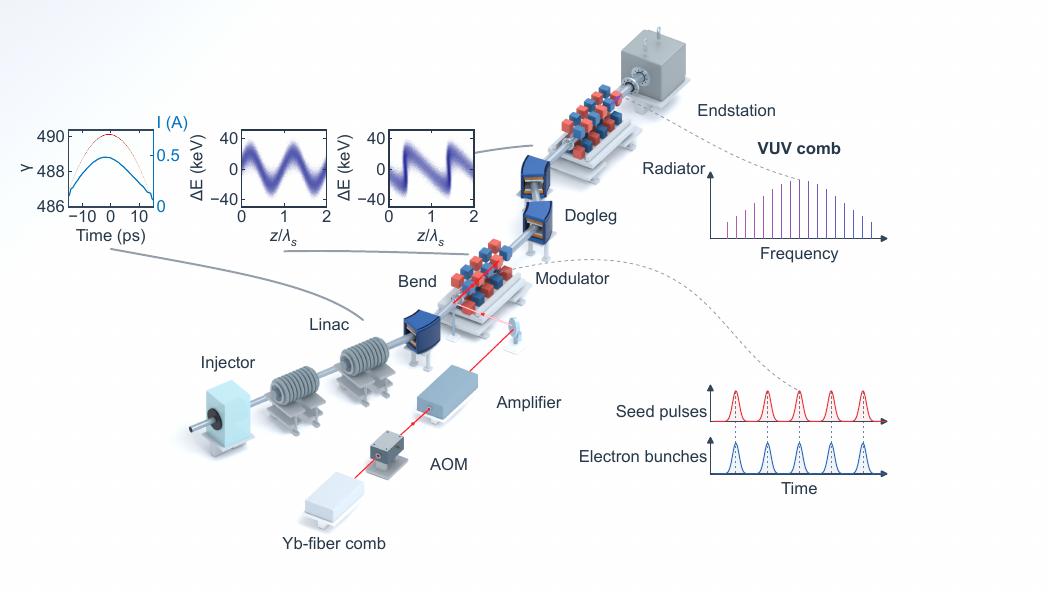}
\caption{Layout of the proposed VUV source. Bunches from
the injector pass through a superconducting linac before
entering the ADM section. The acousto-optic modulator (AOM) shifts the seed
frequency before amplification. The left insets show the energy and current
profiles at the linac exit and the longitudinal phase space at the modulator
exit and radiator entrance. The right insets show how the bunch train is
synchronized with the seed pulses, along with the ideal VUV comb.}
\label{fig:ADM_VUV}
\end{figure*}

\section{Accelerator-Based VUV Comb Source}
\label{sec:source}

The source in Fig.~\ref{fig:ADM_VUV} uses a superconducting radio-frequency
linac to deliver $250\,\mathrm{MeV}$, $10\,\mathrm{pC}$ bunches at
$f_{\rm rep}=1.3\,\mathrm{GHz}$. A synchronized ytterbium laser modulates each
bunch at $1038.7\,\mathrm{nm}$, giving a nominal seventh harmonic at
$148.386\,\mathrm{nm}$. Table~\ref{tab:accelerator} lists the simulation inputs.

\begin{table}[tbp]
\caption{\label{tab:accelerator}
Accelerator and electron-beam parameters used in the simulations. FWHM denotes full width at half maximum and rms denotes root mean square. $N_s$, $N_p$, and $\lambda_u$ denote the number of undulator segments, periods per segment, and period length.}
\begin{ruledtabular}
\begin{tabular}{lc}
Electron beam & \\
\hline
Energy                         & 250 MeV \\
Charge                         & 10 pC \\
Bunch repetition rate          & 1.3 GHz \\
Average beam current           & 13 mA \\
Current profile duration (FWHM) & 21 ps \\
Peak current                   & 0.5 A \\
Normalized emittance (rms)     & 0.15 mm\,mrad \\
Initial slice energy spread (rms) & 5.0 keV \\
\hline
ADM & \\
\hline
Bend angular-dispersion coefficient & 0.1 rad \\
Seed laser wavelength                & 1038.7 nm \\
Seed laser peak power                & 14.5 kW \\
Seed laser intensity duration (FWHM) & 20 ps \\
Seed average power at laser modulator & 400 W \\
Harmonic number $n$                  & 7 \\
Nominal radiation wavelength $\lambda_r$ & 148.386 nm \\
Modulator $N_s \times N_p \times \lambda_u$ & $1 \times 60 \times 50$ mm \\
Radiator $N_s \times N_p \times \lambda_u$ & $1 \times 150 \times 20$ mm \\
\end{tabular}
\end{ruledtabular}
\end{table}

We assume $400\,\mathrm{W}$ average seed power at the modulator,
corresponding to a $14.5\,\mathrm{kW}$ peak power for a Gaussian
$20\,\mathrm{ps}$ intensity profile. An ytterbium fiber amplifier has
delivered comparable average power for a gigahertz pulse train~\cite{fathi2025versatile}.

In the ADM configuration, an upstream bend introduces a horizontal angular
dispersion of $0.1\,\mathrm{rad}$, correlating the electron angle with the
relative energy deviation. The seed laser imposes a periodic energy
modulation in the modulator. The simulated mean modulation amplitude is
about $3.5$ times the initial rms slice energy spread.

The dogleg converts the laser-induced energy modulation into harmonic
microbunching. At the ideal ADM match, the angle- and energy-dependent
longitudinal shifts due to the initial energy deviation cancel to first
order. The bunching factor at harmonic $n$ is then~\cite{FengZhao2017}
\begin{equation}
\label{eq:adm_bunching}
b_n =
J_n\left(
n k_s \xi_D \frac{\Delta\gamma}{\gamma}
\right)
\exp\left[
-\frac{1}{2}
\left(n k_s D_x \sigma_{x'}\right)^2
\right],
\end{equation}
Here $J_n$ is the Bessel function of the first kind, $k_s$ is the seed wavenumber, and $\Delta\gamma/\gamma$ is the relative energy-modulation amplitude. The dogleg coefficients $\xi_D\equiv R_{56}$ and $D_x$ describe longitudinal dispersion and angle-to-longitudinal-position coupling, respectively. The Bessel term describes conversion of the imposed energy modulation, and the exponential describes the reduction in bunching due to the initial horizontal rms angular spread $\sigma_{x'}$.

At the nominal setting used here, particle tracking gives a seventh-harmonic
bunching factor of $|b_7|=0.155$ at the radiator entrance. In the
$3\,\mathrm{m}$ radiator, seventh-harmonic microbunching drives coherent
emission from the transversely matched beam, producing one VUV pulse per
bunch. When the pulses have the same field envelope and a fixed
pulse-to-pulse phase advance, their fields interfere constructively at discrete
frequencies spaced by the bunch repetition rate of $1.3\,\mathrm{GHz}$.

For circular polarization, we model the advanced planar polarized light
emitter II (APPLE-II) radiator as an ideal helical undulator in
GENESIS 1.3~\cite{Reiche1999}. APPLE-II also permits linear
and elliptical polarization~\cite{Hand2016}, offering flexibility to address
magnetic-sublevel transitions and prepare nuclear states~\cite{wense2020}.

\section{Microbunching-Phase Precompensation}
\label{sec:mpp}

\begin{figure}[t]
    \centering
    \includegraphics[width=\columnwidth]{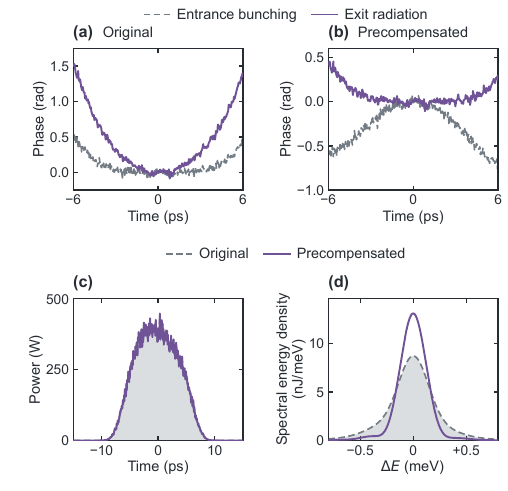}
    \caption{Microbunching-phase precompensation. Constant and linear phase terms are removed in (a,b). The phase panels use different vertical scales. Panels (c,d) use the full transverse field. Panel (d) shows the absolute single-pulse spectral energy density versus $\Delta E=E-E_0$, where $E$ is the photon energy and $E_0$ is the common peak energy.}
    \label{fig:spectral-phase-precompensation}
\end{figure}

At the ideal ADM match, the path-length shifts caused by initial
energy deviations cancel to first order. This suppresses smearing of
the microbunching by the slice energy spread. RF acceleration leaves a
quadratic variation of the slice mean energy along the bunch. The
slices therefore radiate at different detunings from the undulator
resonance, producing the phase curvature in
Fig.~\ref{fig:spectral-phase-precompensation}(a). In MPP, the entrance
microbunching phase shown in Fig.~\ref{fig:spectral-phase-precompensation}(b)
acquires the opposite curvature through a static change to the
dogleg's longitudinal dispersion $R_{56}$.

A dispersion change $\Delta R_{56}$ shifts a slice's path length by
$\Delta R_{56}\bar{\delta}(t)$ and its bunching phase by
$-k_r\Delta R_{56}\bar{\delta}(t)$. The slice mean relative energy
offset is $\bar{\delta}(t)$ and the radiation wavenumber is
$k_r=2\pi/\lambda_r$. In a one-dimensional low-gain model, near-resonant
emission over $N_u$ periods adds a phase $-2\pi N_u\bar{\delta}(t)$ to
first order. In the reference setting, the entrance bunching phase
has a nearly zero quadratic term. The two terms proportional to
$\bar{\delta}(t)$ cancel when
\begin{equation}
\Delta R_{56}\simeq -N_u\lambda_r .
\label{eq:mpp_phase_balance}
\end{equation}
For the 150-period radiator, the shift is about $-22\,\mu\mathrm{m}$,
roughly $1\%$ of the reference $|R_{56}|$. To first order, a small
achromatic chicane between the dogleg and the radiator could trim the
total $R_{56}$ independently of the angle-to-longitudinal-position
coupling $D_x$~\cite{Li2022ImprovedADM}. The three-dimensional
comparison holds the incoming beam, seed, and radiator fixed.

As shown in Fig.~\ref{fig:spectral-phase-precompensation}(c), MPP
changes the pulse energy and temporal power profile at the radiator
exit very little.
The single-pulse spectral full width at half maximum falls from
$0.38$ to $0.28\,\mathrm{meV}$, and the interval containing $90\%$
of the spectral energy narrows from $0.84$ to $0.45\,\mathrm{meV}$.
The peak of the transversely integrated spectral energy density in
Fig.~\ref{fig:spectral-phase-precompensation}(d) rises from $8.72$ to
$13.10\,\mathrm{nJ/meV}$, an increase of about $50\%$. Relative to a
flat-phase reference with the corresponding field amplitude in space
and time, the single-pulse FWHM falls from $1.47$ to $1.10$ times the
reference width.

We evaluate the phase at line $b$ of the $^{229}$Th:CaF$_2$ crystal
TS1. Its measured frequency is $\nu_b\simeq2020.407299\,\mathrm{THz}$
at $301.4\,\mathrm{K}$~\cite{Huang2026}. We project the exit Fourier
field at $\nu_b$ onto a fixed transverse mode to obtain the complex
amplitude $a$, with phase $\phi=\arg a$ referenced to the seed and
laboratory time. A centroid-energy shift $\Delta E_b$ changes
$\bar{\delta}$ by $\Delta E_b/E_{b0}$, where $E_{b0}$ is the reference
centroid energy. The added transport phase gives
$S_{\rm MPP}-S_0\simeq-k_r\Delta R_{56}/E_{b0}$, where $S=d\phi/dE_b$
and $S_0$ is the slope before compensation. In the energy scan, MPP
reduces $|S|$ by about $85\%$. With combined beam and seed jitter, MPP
substantially reduces the simulated shot-to-shot phase spread at
$\nu_b$. The coherent component accounts for about $99\%$ of the mean
power in the selected transverse mode.

With the seed phase fixed, the tooth near $\nu_b$ reconstructed from
200 simulated pulses has a FWHM of about $5.76\,\mathrm{MHz}$,
consistent with the Fourier width of an ideal equal-amplitude,
phase-locked train over the same $154\,\mathrm{ns}$ window. A single
static dispersion adjustment thus narrows the single-pulse spectrum
and makes the phase at the nuclear frequency more reproducible at the
radiator exit.

\section{Nuclear Spectroscopy with the VUV Comb}
\label{sec:nuclear_spectroscopy}
\label{sec:comb}

\begin{figure}[h!]
\centering
\includegraphics[width=\columnwidth]{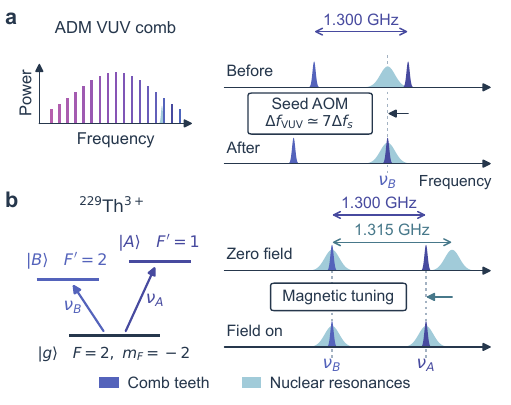}
\caption{Comb tuning for two-tooth resonance in $^{229}\mathrm{Th}^{3+}$.
(a) A seed AOM shifts the comb to align a tooth with $\nu_B$.
(b) The bias field tunes $\nu_A-\nu_B$ to the $1.300\,\mathrm{GHz}$
tooth spacing. Spectra are schematic.}
\label{fig:comb_tuning_dual}
\end{figure}

At the fixed $1.300\,\mathrm{GHz}$ tooth spacing, the seed AOM in
Fig.~\ref{fig:comb_tuning_dual}(a) translates the VUV comb by seven
times the seed shift under ideal seventh-harmonic transfer. For an
ideal coherent train, the simulated precompensated field gives $21.8\,\mathrm{mW}$ in the
tooth aligned to the crystal $b$ line out of $5.32\,\mathrm{W}$ total
average power at the radiator exit ($0.41\%$).

\begin{figure}[h!]
    \centering
    \includegraphics[width=\columnwidth]{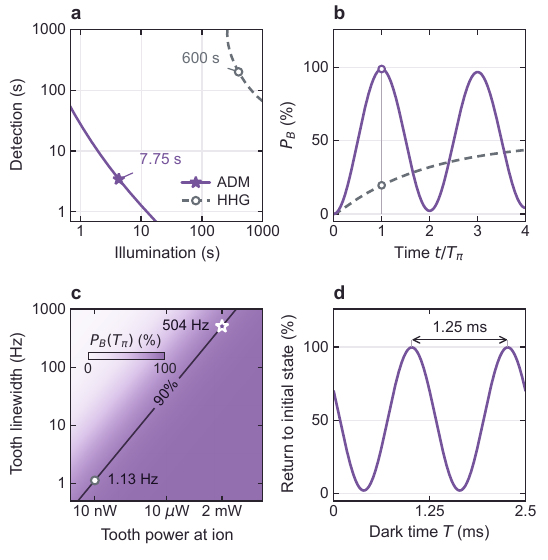}
    \caption{Modeled nuclear spectroscopy.
    (a) Crystal fluorescence readout at equal SNR, with total times marked.
    (b) Single-tooth excitation at a common $49\,\mathrm{Hz}$ Lorentzian
    linewidth, with time scaled by $T_\pi=1/(2f_{\mathrm{R}})$ for each source.
    (c) Linewidth limits for $90\%$ excitation at $T_\pi$ (markers).
    Both panels use a uniform $3\,\mu\mathrm{m}$-diameter spot at the ion.
    (d) Ramsey return for equal and opposite tooth detunings at fixed
    relative readout phase ($0.183\,\mathrm{ms}$ pulses, $1\,\mathrm{Hz}$
    common optical linewidth).}
    \label{fig:nuclear_spectroscopy}
\end{figure}

Scanning a tooth across the crystal $b$ line locates the clock
resonance from its fluorescence. More power in the resonant tooth
excites the nuclei faster and shortens the acquisition. We compare
the combs at $200\,\mu\mathrm{W}$ total incident power and an assumed
common $490\,\mathrm{Hz}$ Lorentzian tooth linewidth~\cite{wense2020}.
With spectrally uniform transport, the model gives $819\,\mathrm{nW}$
in the resonant ADM tooth at the sample, compared with the reported
$1\,\mathrm{nW}$ estimate for high-harmonic generation
(HHG)~\cite{Zhang2024}. Using the measured TS1 $b$-line core and the lifetime of the
unresolved feature~\cite{Huang2026}, the model in
Fig.~\ref{fig:nuclear_spectroscopy}(a) gives acquisition times of
$7.75\,\mathrm{s}$ for ADM and about $600\,\mathrm{s}$ for HHG
at the same on-resonance SNR.

At the same focus, $2\,\mathrm{mW}$ in the ADM tooth at the ion gives
a resonant Rabi frequency $f_{\mathrm{R}}$ more than two orders of
magnitude above that of the HHG reference~\cite{wense2020}. We use the
inferred nuclear radiative rate~\cite{Tiedau2024} to model the $|g\rangle\to|B\rangle$ M1 transition
of trapped $^{229}\mathrm{Th}^{3+}$. The faster drive leaves less time for
phase diffusion, preserving the Rabi oscillations in
Fig.~\ref{fig:nuclear_spectroscopy}(b) at the common linewidth. As shown in
Fig.~\ref{fig:nuclear_spectroscopy}(c), ADM still reaches $90\%$
excitation at $T_\pi$ with a tooth linewidth of about $504\,\mathrm{Hz}$.

We estimate the isomeric $5f_{5/2}$ hyperfine constants by scaling the
measured ground-state constants by the isomer-to-ground ratios obtained
from $6d_{3/2}$ spectroscopy~\cite{Yamaguchi2024,Campbell2011ThHyperfine,Zitzer2025ThHyperfine}.
This gives a zero-field separation $\nu_A-\nu_B$ of about
$1.315\,\mathrm{GHz}$ for the two $\sigma^+$ transitions from $|g\rangle$.
In Fig.~\ref{fig:comb_tuning_dual}(b), a bias field tunes the separation
to the $1.300\,\mathrm{GHz}$ tooth spacing through the differential
Zeeman shift. Shifting the comb as a whole then aligns the lower and
higher teeth with $\nu_B$ and $\nu_A$, respectively.

Ramsey readout compares the nuclear splitting with the beat of the
same two teeth at $f_{\mathrm{rep}}$. The first pulse prepares a superposition of
$|A\rangle$ and $|B\rangle$. During the dark time $T$, the relative
phase of the two states slips against the comb beat if the nuclear
splitting differs from the tooth spacing. A second pulse from the same
pair makes the two paths back to $|g\rangle$ interfere, turning the
phase difference into a variation of the return probability $P_g(T)$.
For a fixed relative readout phase and after averaging over the common
readout phase, the fringe frequency gives the magnitude of the mismatch
between the nuclear splitting and the known tooth spacing,

\begingroup
\abovedisplayskip=\dimexpr\abovedisplayskip\relax
\abovedisplayshortskip=\dimexpr\abovedisplayshortskip\relax
\belowdisplayskip=\dimexpr\belowdisplayskip\relax
\belowdisplayshortskip=\dimexpr\belowdisplayshortskip\relax
\begin{equation}
f_{\mathrm{Ramsey}}=\big|(\nu_A-\nu_B)-f_{\mathrm{rep}}\big|.
\label{eq:ramsey_difference}
\end{equation}
\endgroup
For the illustrative detunings in Fig.~\ref{fig:nuclear_spectroscopy}(d),
increasing the dark time by $1.25\,\mathrm{ms}$ spans one full cycle of
the calculated return probability.

\section{Conclusion}

In summary, for a coherent pulse train, simulations predict milliwatt-scale
power in a resonant tooth at the exit of the accelerator-based VUV source.
Models for $^{229}$Th predict faster crystal fluorescence readout and
submillisecond nuclear $\pi$ pulses in trapped ions. Two adjacent teeth
of the same comb could prepare a superposition of two isomer hyperfine
levels from a common ground state and probe it by Ramsey interferometry.
These capabilities could advance optical nuclear clocks and support the
development of quantum memories that encode information in superpositions
of nuclear states.

\begin{acknowledgments}
This work was supported by the National Natural Science Foundation of China (12625511, 12435011), CAS Project for Young Scientists in Basic Research (YSBR-115), and Shanghai Municipal Science and Technology Major Project.
\end{acknowledgments}

\let\originalbibfont\bibfont
\def\bibfont{\originalbibfont\interlinepenalty=10000}
\makeatletter
\newcount\ADMBibNameCount
\def\ADMBibAuthorField{author}
\def\ADMBibInfoToken{\bibinfo}
\def\ADMBibPrintFirstThree#1\bibinfo#2\bibinfo#3\bibinfo#4\bibinfo#5\ADMBibEnd{%
  #1\bibinfo#2\bibinfo#3\bibinfo#4\emph{et~al.},\ %
}
\protected\def\bibfield#1#2{%
  \begingroup
  \def\ADMBibFieldName{#1}%
  \ifx\ADMBibFieldName\ADMBibAuthorField
    \ADMBibNameCount=0\relax
    \@tfor\ADMBibToken:=#2\do{%
      \ifx\ADMBibToken\ADMBibInfoToken
        \advance\ADMBibNameCount by 1\relax
      \fi
    }%
    \ifnum\ADMBibNameCount>6\relax
      \ADMBibPrintFirstThree#2\ADMBibEnd
    \else
      #2%
    \fi
  \else
    #2%
  \fi
  \endgroup
}
\makeatother
\bibliography{sample}

\end{document}